\documentclass[12pt]{iopart}
\usepackage{cite,here,colordvi}
\usepackage{graphicx}

\begin{document}
\title[Relativistic precession]
{Relativistic precession and spin dynamics of an elliptic Rydberg wave packet}
\author{P Rozmej\dag, M Turek\ddag, R Arvieu$^\spadesuit$,  
~and I S Averbukh$^\clubsuit$
\footnote[5]{Research supported by KBN grants No.~5~P03B~010~20 and 
5~P03B~104~21}} 

\address{\dag ~Institute of Physics, University of Zielona G\'ora,  
65-246 Zielona G\'ora,  Poland \\ {~p.rozmej@im.pz.zgora.pl}~}
\address{\ddag ~Institute of Physics,  Maria Curie-Sk\l odowska University,   
20-031 Lublin,   Poland\\ {~mturek@kft.umcs.lublin.pl}~}
\address{$^\spadesuit$ Institut des Sciences Nucl\'eaires,   
F 38026 Grenoble Cedex,  France\\ {~arvieu@in2p3.fr}~}
\address{$^\clubsuit$ The Weizmann Institute of Science
76100 Rehovot,  Israel \\{~ilya.averbukh@weizmann.ac.il}~}

\begin{abstract}
Time evolution of wave packets built from the eigenstates of the Dirac equation
for a hydrogenic system is considered. We investigate the space and spin motion
of wave packets which, in the non-relativistic limit, are stationary states with
a probability density distributed uniformly along the classical, elliptical orbit
(elliptic WP). We show that the precession of such a WP, due to relativistic 
corrections to the energy eigenvalues, is strongly correlated with the spin motion.
We show also that the motion is universal for all hydrogenic systems with an
arbitrary value of the atomic number $Z$.  
\end{abstract}

The detailed study of the time evolution of quantum 
wave packets (WPs) in simple atomic or
molecular systems has been the object of growing attention for more
than ten years \cite{r1}. Most of the previous theoretical studies were done   
in non-relativistic framework.  In the field of relativistic quantum
mechanics most of the efforts have been focused on the problem of the
interaction between the atoms and a mixture of static fields with, most of the
time, intense laser fields \cite{r2,r3,r4,r5,r6,r7,r8,r9,r10,r11}.  
Under these conditions the use of a
relativistic theory is fully justified since the external field is then
able to bring considerable energy to the WP.  For isolated atoms, however,
the use of relativistic dynamics is more questionable, if the WP is
followed or observed only during a short period of time.  In ref. \cite{r12}
relativistic  wave packets, corresponding to circular orbits,
have been constructed for hydrogenic atoms
with a large $Z$, and propagated over time according to the Dirac equation.  
Particular attention was paid to the spin collapse event, {\em i.e.} 
 to the
maximum entanglement, in the course of time, of the spin degree of freedom
with the spatial ones.  This phenomenon was first shown to take place for a WP in
a harmonic oscillator with a spin-orbit force \cite{r13}, where it is
periodic.  For this reason it has been called the {\em spin-orbit pendulum}. 
In the
Dirac equation with a Coulomb potential, it is produced by the built-in
spin-orbit force and it is not periodic.  The time scale where this effect
can manifest itself was discussed in ref. \cite{r12}, as a function of the charge $Z$ 
of the atom and the average principal quantum number $N$ of the WP.  
This phenomenon
is expected to take place on a longer time scale like the other time dependent
quantum effects of spreading and of fractional revivals \cite{r14}.  
We intend
below to complement this work by studying the possible relativistic
precession of elliptic WPs and by comparing this precession to the spin
motion. A preliminary version of the present work was already presented 
at the ECAMP VII conference \cite{ecamp}.
   
   There are two possible ways to build up a WP 'on top of a classical
elliptic orbit' in non-relativistic mechanics.  One of them is by
kicking properly a well designed WP, for example a Gaussian WP, 
that is then evolved
in time by the free Hamiltonian until it spreads above an average
classical ellipsis.  This method is not very easy to apply and its
main inconvenience is to produce an internal motion within the ellipsis
that is able to blur the precession.  Therefore, we have preferred a second
method, much simpler and even more elegant, which consists of using the
coherent WPs of ref.~\cite{r15}, which are stationary states of the 
non-relativistic Coulomb problem.  
The space probability density of these states was
indeed shown to be highly concentrated around a classical Bohr-Sommerfeld
ellipsis.  If these states can be adapted to relativistic dynamics their
time evolution will doubtlessly be due to relativistic effects, {\em i.e.}  
due to the fine structure that will act as a perturbation.  
  Let us first show how to adapt these states to a relativistic
theory. We want the spatial part of the large components of the new WP to tend
(in the non-relativistic limit) 
toward the state $|n \gamma\rangle $ of \cite{r15} which is defined as  
\begin{eqnarray}\label{e1}
\fl
|n \gamma\rangle  =  \sum_{l,m} (-1)^{(l+m)/2} \frac{2^{n-l-1}(n-1)!}
 {[(l-m)/2]![(l+m)/2]!} \left[ \frac{(l+m)!(l-m)!(2l+1)}{(n-l-1)!(n+l)!} 
 \right]^{1/2} \nonumber \\
  \times \left(\sin \frac{\gamma}{2}\right)^{n-m-1}
 \left(\cos \frac{\gamma}{2}\right)^{n+m-1}\, |n,l,m\rangle 
 \;\; = \;\; \sum_{lm} w_{lm}^{(n)}\,|n,l,m\rangle  \; . 
\end{eqnarray}

 The probability density $\langle n\gamma |n \gamma\rangle$ 
 is fairly localized onto a Bohr orbit with
eccentricity $\epsilon=\sin \gamma$ and the average angular momentum is 

\begin{equation}\label{e2}
  l_{av} = (n-1)\, \cos{\gamma} \;,                  
\end{equation}
where  $n$ is the principal quantum number of the orbitals 
 $|n,l,m\rangle$ which are admixed in~(\ref{e1}).  
The sum contains $n(n+1)/2$ values of $m$ with $l+m$ even.  The
contribution of states with $m<l$ decreases very rapidly with $m$ (more than
one order of magnitude for each 2 units of $m$ as shown in fig.~1). 
The dominant weights are those with $m=l$ and their distribution is nearly
Gaussian. 
The relative contribution of states with $m<l$ increases, however,
for larger values of the eccentricity parameter $\epsilon$.  
The larger admixture of these states causes much faster 
daphasing of the WP. Therefore for illustration of typical precession phenomena
he have chosen the case $\epsilon=0.4$.

    In order to study the entanglement of the spin degrees of freedom with
the orbital ones, it is natural in a non-relativistic theory to start
from a product state of an arbitrary spinor 
$\left( \begin{array}{c}a\\ b \end{array}\right)$
 with the state $|n \gamma\rangle$.  
\begin{equation}\label{e3}
  |\Psi_{nr}\rangle  =  |n \gamma\rangle \,\left( \begin{array}{c}
  a\\ b \end{array}\right)  
   =     \sum_{lm} w_{lm}^{(n)}\,|n,l,m\rangle \,\left( \begin{array}{c}
  a\\ b \end{array}\right) \,.           
\end{equation}
It may be expanded in eigenstates of total angular momentum 
$|n,l,j=l+s,m_j\rangle$  with
$m_j=m+1/2$ or $m-1/2$ and $s=+1/2$ or $-1/2$. 

 \begin{eqnarray}\label{e5}
 \fl
|\Psi_{nr}\rangle = \sum_{lm} \, w_{lm}^{(n)} \,
\left\{ \,
 a \, \left( \,
    \sqrt{\frac{l+1+m}{2l+1}} \, | n, l, j_>, m+1/2 \rangle
  - \sqrt{\frac{l-  m}{2l+1}} \, | n, l, j_<, m+1/2 \rangle \,
          \right) \right. \\
+\left. 
 b \, \left( \,
    \sqrt{\frac{l+1-m}{2l+1}} \, | n, l, j_>, m-1/2 \rangle
  + \sqrt{\frac{l+  m}{2l+1}} \, | n, l, j_<, m-1/2 \rangle \,
          \right) \right\} \nonumber  \,. 	  
\end{eqnarray}	                        
 In a similar way as in ref.~\cite{r12}, for a circular WP, 
 the state (\ref{e5}) is
transformed into a four component relativistic state $\Psi_r$ by replacing
in (\ref{e5}) the non-relativistic states $|n,l,j,m_j\rangle$ 
by the eigenstates of the Dirac
equation with the same quantum numbers.  In this manner the WP gets small
components in the most natural way.  The radial parts of the large and of
the small components are taken obviously as different ones.  The
probability density of the relativistic WP built in this way is represented in
 fig.~2.  

  The time evolution of the WP is produced by introducing the phase
factors $\exp{(-iE_l^+\,t)}$ and $\exp{(-iE_l^-\,t)}$
as coefficients of states with $j=l+1/2$ and $j=l-1/2$
with their corresponding eigenvalues. The four components $c_i(t)$ $i=1,\ldots, 4$ 
 of the WP at time $t$ are given in the \ref{aA}
 with  $\Psi_r$ defined as 
\begin{equation}\label{e6}
 |\Psi_r(t)\rangle = \left( \begin{array}{c} |c_1(t)\rangle \\
 |c_2(t)\rangle \\|c_3(t)\rangle \\|c_4(t)\rangle  \end{array} 
 \right) \;.
\end{equation}
  
   It should be stressed that $\Psi_r$ for $t=0$ is not any more a product
state of the form of eq.~(\ref{e3}) due to the built-in 
entanglement contained in the solutions of the Dirac equation. 
However, since the small components of $\Psi_r$ 
are indeed very small, this defect has no important effect on the
magnitude of the initial spin. 
Since the spin-orbit coupling effect, {\em i.e.}~spin-orbit
pendulum \cite{r13}, manifests itself more efficiently if $s$ and $l$ are
perpendicular to each other, we choose for our discussion:
\begin{equation}\label{e7}
 a = b =\frac{1}{\sqrt{2}} \quad\quad \mbox{\em i.e.}
 \quad\quad  \langle s_x \rangle_{nr} = \frac{1}{2} \;.
\end{equation}
                         
Precession of quantum elliptical states in the Coulomb field has already been 
considered in ref.~\cite{crawford}, starting also from the same coherent state
as done here. However, the precession was studied only in non-relativistic quantum
dynamics as a perturbation effect and no treatment of the spin was attempted. 
				 				 
 Let us finally discuss the time units relevant to our problem. To the
lowest order in terms of the fine structure constant, the energy of an
eigenstate $|n,l,j=l+s\rangle$ of the Dirac equation can be written (in a.u.) 
for a hydrogenic atom of charge $Z$ as 
\begin{equation}\label{e8}
 E_{nlj} = \overline{E}_n-\frac{1}{2} \frac{Z^4\alpha^2}{n^3(l+s+1/2)}
\end{equation}
  with $s=+1/2$ or $-1/2$. 
 The constant energy $\overline{E}_n$ produces no effect on WP, since it
depends only on $n$. Let us define an average time unit $T_p$ ($p$ for
precession): 
\begin{eqnarray}\label{e9}
 T_p & = & \frac{2\pi}{\langle \frac{dE_{nlj}}{dl} \rangle} \\
     & = & \frac{4\pi n^3}{Z^4\alpha^2} \langle (l+s+\frac{1}{2})^2\rangle
     \approx \frac{4\pi n^3\,l_{av}^2}{Z^4\alpha^2} \label{e10}\\
     & = & \frac{2 \,l_{av}^2\,T_K}{(Z\alpha)^2}\;.\label{e10a}
\end{eqnarray}
Brackets $\langle\rangle$ in (\ref{e9})-(\ref{e10}) denote average values,
$l_{av}$ is given by (\ref{e2}). 
For $n=50$, $\epsilon=0.4$ precession time 
$T_p$ ranges from $1.96\cdot 10^{-11}$s for $Z=92$ to $1.4\cdot 10^{-3}$s 
for  $Z=1$. $T_K$ in  (\ref{e10a}) denotes the Kepler period.

It is necessary to compare $T_p$ to the radiative lifetime 
$T_{n,l}^{\mbox{\tiny rad}}$  of hydrogenic levels.
We will use the estimation of $T_{n,l}^{\mbox{\tiny rad}}$ for
a single $n,l$ state given in \cite{chang}. In atomic units it is 
\begin{equation} \label{trad}
 T_{n,l}^{\mbox{\tiny rad}} = \frac{3}{2\,\alpha^3\,Z^4}\,n^3
  \left(l+\frac{1}{2}\right)^2\;,
\end{equation}
which was found to agree within 10\% with experimental data. 
One obtains the universal ratio:
\begin{equation} \label{tptr}
 \frac{T_p}{T_{n,l}^{\mbox{\tiny rad}}} = \frac{8\pi\,\alpha}{3} \approx 0.061
\end{equation}
which guarantees that the precession of the wave packet takes place long enough 
time before even a single photon is emitted. 
The occurrence of $\alpha$ in this ratio is understandable,
since $T_p$ is a classical-like quantity, 
while $T_{n,l}^{\mbox{\tiny rad}}$ is proportional to $1/\hbar$, because it
can be expressed as a ratio of a typical energy of the emitted photon to 
the classical radiation rate corresponding to the orbital motion.

When $t=T_p$ the linear terms in the expansion of $E_{nlj}$ in the power of $l$
contribute on the average to $2\pi$ in the exponential factors and the WP
is expected to restore. However the terms of the higher order 
dephase differently the various partial waves, and this leads to a
spreading of the WP near the initial shape \cite{r14}. 
See the discussion on the magnitude of these terms in \ref{aB}.
Expression (10) or (11) (with the
Kepler period $T_K=2\pi n^3/Z^2$) is also recognized as the 
classical precession time
in the relativistic Coulomb problem \cite{r16}. 

      Let us note that $dE_{nlj}/dl$
is also (to the first order) the energy
difference between two spin-orbit partners. 
Therefore the precession time can also be
interpreted as the recurrence time of the spin. Hence we should
observe a strong correlation between the spatial motion of the density: 
the precession, and the spin motion. 

 Let us discuss now the dynamics dependence on the atomic number, $Z$.
Formula (\ref{e10}) suggests that the relativistic effects under consideration
in this article, depend crucially on $Z$.
For sure the highest possible $Z$ are indeed
required to lower $T_p$ as much as possible. It is interesting to stress
nevertheless, that within a very good approximation, a scaling of $Z$ is
possible which leads to the universal behaviour of the wave packet (\ref{e1}).

   First of all since $E$, approximated by eq.~(\ref{e8}),
scales as $Z^4$, {\em i.e.} as $T_p^{-1}$, the variable  $Z$ disappears 
from $Et$ if we use the reduced time $t/T_p$.
 The autocorrelation function (\ref{ac}) which is 
expressed only in terms of $Et$ is the simplest quantity which has a 
universal behaviour, provided the same $n$
and $\epsilon$ are used for all values of $Z$.

   The other quantities, like the probability density and the the spin
expectation values depend on the radial wave functions and radial
integrals. In a non-relativistic theory scaling of the radial wave function
is elementary, it is obtained by dividing the radial wave fuction by
$Z^{3/2}$ and multiplying the radial variable by $Z$. This is an exact
property. The ratio small/large components of the relativistic solutions
scale as $\sqrt{(1-E)/(1+E)}$, {\em i.e.}~roughly like 
$Z^2$ and the radial wave functions 
depend also in a more complicated manner on $Z$. 
Nevertheless, on the whole as seen in fig.~5 below, the small components 
contribute a very small part of the probability density even
for $Z=92$ (see also fig.~1 of ref.~\cite{r12}). 
The scaling of the probability
density is therefore entirely governed by the large components {\it i.e.} 
by the non-relativistic theory.

   In a similar way we have checked that the radial integrals which
contribute to the spin expectation values have the same properties:
the integrals $G_+, G_-$, and $G_{+-}$ are equal to 1, within less than 
10$^{-5}$ and the other $F_+ ,F_- \dots$ contribute in a very small manner,
also of the order of 10$^{-5}$.

   Therefore, the universal behaviour of the wave packet is justified and
except for fig.~2 no value of $Z$ is given. For longer times, the
energy factors omitted in eq.~(\ref{e8}) which involve higher powers of $Z$, 
play a role and produce a genuine $Z$ dependence. Those effects will not be
discussed here.

   The probability density of the wave packet with $n=50, \epsilon=0.4$ and
$a=b=1/\sqrt{2}$, with $Z=92$, is shown for a set of 
times up to $t=T_p$  in fig.~3 and  fig.~4.
The part of the density coming from the small components,  shown in
fig.~5, is also concentrated on top of an ellipsis but its
magnitude is a thousands times smaller than the total density. For $t<T_p/4$
the density precesses as described classically with a small
dispersion. However the spreading takes place very rapidly for larger $t$ and
is quite extended for $t=T_p$. 

    The precession of the ellipsis, the recurrence and spreading can be
seen in a more condensed manner in the autocorrelation function
represented in fig.~6 for three different spin preparations 
(spin up, spin down and $a=b=1/\sqrt{2}$).
For small $t/T_p$ the WP becomes almost orthogonal 
to its initial parent, for $t=T_p$ a recurrence occurs but the overlap is only
0.7. Another peak occurs at $t=2T_p$ but higher frequencies become important
and spread the recurrence. For $t>3T_p$ these higher frequencies play a
dominant role.  An example of an approximate revival for $t=22.6\,T_p$ is 
displayed in fig.~7. 
Fractional revivals \cite{r14} can also be seen to some extent. Two examples 
of 1/3 and 1/2 revivals are presented in fig.~8.
    
    The rough independence of the autocorrelation function on the spin
preparation requires some explanation. We can approximate this function by
\begin{equation}\label{ac1}  
 \langle \Psi_{r}(0)|\Psi_{r}(t) \rangle \approx a^2 \sum_l   
   w_{ll}^2 \exp{(-iE^+_l t)}+b^2 \sum_l  w_{ll}^2 \exp{(-iE^-_l t)} \;, 
\end{equation}  
where we have neglected the terms with very small weights 
$w^2_{l,m\neq l}$. The contribution of the 
 two sums above is
almost the same because $w_{ll}$ has a smooth variation with $l$ on the one hand,
and because of the exact equality   
          $ E_l^-=E_{l-1}^+ $ on the other hand.      
	    
Finally the time evolution of the spin averages is presented in fig.~9. 
This figure exhibits what we can call the relativistic spin-orbit
pendulum. Although the wave packet is not prepared initially 
in a pure state of spin, 
the impurity is very small (for $a=b=1/\sqrt{2}$ one has
$\langle \sigma_x\rangle \approx 0.9997$). 
As time goes on the spin stays very nearly in the $Oxy$ plane
and rotates around $Oz$ with period $T_p$. Its magnitude slowly decreases,
however, and for $5<t/T_p<10$ the spin is amost totally entangled with the
orbital motion, since the average of its three projections are almost zero. 
During a period of time that last for about $5T_p$ the angular momentum of the
spin is transferred to the orbital motion and therefore the mean
trajectory is not planar anymore \cite{r13}. Since the orbital angular momentum 
of $40\hbar$ is much higher than $\hbar/2$ this geometrical effect can hardly
be represented graphically. From $t\approx 15T_p$ a revival of spin occurs. 
The spin rotates and increases its magnitude at the same time. This event 
stays even longer than initially. However the recurrence is only partial.

In conclusion we can say, that for not too long time, the precession and 
spin motion of the WPs are fairly well described by the following
approximation: non-relativistic wave functions of the form (\ref{e3}) and
relativistic energy eigenvalues. From this point of view the full,
computationally very demanding, relativistic approach is unnecessary.
However, this conclusion may be formulated not a priori but only a posteriori.

We would like to stress the richness of the dynamics just described. 
Indeed, in addition to the relativistic precession of the ellipsis we have
obtained for longer times its fractional revivals. During the evolution the spin
of the electron is entangled with its orbital motion to various degrees. 
All this agree completely with previous results \cite{r14} and \cite{r13}. 
However, it was obtained here in a full relativistic calculation. Since we have
been able to scale the atomic number $Z$ we have given a universal behaviour to
our WP. It is, however, clear that this scaling is destroyed in real atoms in a
more realistic theory which would take into account quantum defects. Their
inclusion would also distort the dynamics, for example it would change the
precession time, in a way that is out of reach of our simple theory. As far as
purely relativistic effects are concerned, like the importance of the small
components or the zitterbewegung, we have found them negligible in the Coulomb
problem, in contrast to the Dirac oscillator \cite{r19} in which they play a
major role.

\ack
The authors would like to express their gratitude to the two referees
who attracted their attention to the interplay between relativistic
precession and radiative processes (eq. (12)) in the Coulomb problem.
I.S. Averbukh thanks also C.R. Stroud, Jr. for useful discussions.

\appendix
\section{Details of the calculations}\label{aA}

Replacing a non-relativistic wave function in (\ref{e5})
by the eigenstates of the Dirac equation, one obtains for $\Psi_r(t)$:
\begin{eqnarray}\label{a1}
\fl
\Psi_r(t)=\sum_{lm} \, w_{lm}^{(n)} & & 
\left\{  a \sqrt{\frac{l+1+m}{2l+1}}
 \pmatrix{ i g_{n'_{+}} \Omega_{l, j_>, m+1/2} \cr
 - f_{n'_{+}} \Omega_{l+1, j_>, m+1/2}        }
       \exp{(-i E^+_lt)}\right.\nonumber
\\ 
  & & \left. + a \sqrt{\frac{l-m}{2l+1}}
\pmatrix{ i g_{n'_{-}} \Omega_{l, j_<, m+1/2} \cr
 - f_{n'_{-}} \Omega_{l-1, j_<, m+1/2}        } 
       \exp{(-i E^-_lt)}\right.
\\
& &\left.  
+ b \sqrt{\frac{l+1-m}{2l+1}}
 \pmatrix{ i g_{n'_{+}} \Omega_{l, j_>, m-1/2} \cr
 - f_{n'_{+}} \Omega_{l+1, j_>, m-1/2}        }
       \exp{(-i E^+_lt)}\right.\nonumber
\\ 
  & & \left. - b \sqrt{\frac{l+m}{2l+1}}
\pmatrix{ i g_{n'_{-}} \Omega_{l, j_<, m-1/2} \cr
 - f_{n'_{-}} \Omega_{l-1, j_<, m-1/2}        } 
       \exp{(-i E^-_lt)}\nonumber\right\} \;.
\end{eqnarray}
We have used the notations of \cite{r12}: $g(r)$ and $f(r)$ are the radial parts
of the large and small components associated with the quantum numbers 
$n'=n-(j+1/2)$. These functions are multiplied by the spherical tensors 
$\Omega_{l, j, m_j}$ which are defined by eq.~4a and 4b of \cite{r12}. 
The energy of the spin orbit partners of a given value of $l$ is denoted 
by $E_l^+$ if $j=l+1/2$ and $E_l^-$ if $j=l-1/2$, respectively.

For numerical calculations it is convenient to rewrite components of (\ref{a1})
in the following form:
\begin{eqnarray}
\fl
|c_1(t)\rangle = i \sum_{l} \left\{ g_{n'_+} \exp{(-iE^+_lt)} \sum_{m} \, 
  w_{lm}^{(n)} \left( a \frac{l+1+m}{2l+1}|l,m\rangle 
  \right.\right. \nonumber
  \\  \left.\quad\quad\quad\quad\quad\quad\quad\quad\quad
  +b \frac{\sqrt{(l+1-m)(l+m)}}{2l+1} |l,m-1\rangle \right)\\
  + \, g_{n'_-} \exp{(-iE^-_lt)}\sum_{m} \, w_{lm}^{(n)} \left( 
  a \frac{l-m}{2l+1}|l,m\rangle 
  \right.\nonumber
  \\  \left.\left.\quad\quad\quad\quad\quad\quad\quad\quad\quad
  -b \frac{\sqrt{(l+1-m)(l+m)}}{2l+1} |l,m-1\rangle \right)\right\}
  \nonumber \;,
\end{eqnarray}

\begin{eqnarray}
\fl
|c_2(t)\rangle = i \sum_{l} \left\{ g_{n'_+} \exp{(-iE^+_lt)} \sum_{m} \, 
  w_{lm}^{(n)} \left( b \frac{l+1-m}{2l+1}| l,m \rangle
  \right.\right. \nonumber
  \\  \left.\quad\quad\quad\quad\quad\quad\quad\quad\quad
  +a \frac{\sqrt{(l+1+m)(l-m)}}{2l+1} |l,m+1\rangle \right) \\
   + \, g_{n'_-} \exp{(-iE^-_lt)}\sum_{m} \, w_{lm}^{(n)} \left( 
  b \frac{l+m}{2l+1}| l,m \rangle 
  \right.\nonumber
  \\  \left.\left.\quad\quad\quad\quad\quad\quad\quad\quad\quad
  -a \frac{\sqrt{(l+1+m)(l-m)}}{2l+1} |l,m+1\rangle \right)\right\}
  \nonumber \;,
\end{eqnarray}

\begin{eqnarray}
\fl
|c_3(t)\rangle  = \sum_{l} \left\{ f_{n'_+} \exp{(-iE^+_lt)}  \sum_{m} \,  
  w_{lm}^{(n)} \left(
  a \sqrt{\frac{(l+1+m)(l+1-m)}{(2l+1)(2l+3)}}|l+1, m \rangle 
  \right.\right. \nonumber
  \\  \left.\quad\quad\quad\quad\quad\quad\quad\quad\quad
  + b \sqrt{\frac{(l+1-m)(l+2-m)}{(2l+1)(2l+3)}}|l+1, m-1\rangle
  \right) \\ 
    + \,f_{n'_-} \,\exp{(-iE^-_lt)}  \sum_{m} \,  w_{lm}^{(n)} \left( 
  a \sqrt{\frac{(l+m)(l-m)}{(2l+1)(2l-1)}}|l-1,m\rangle 
  \right.\nonumber
  \\  \left.\left.\quad\quad\quad\quad\quad\quad\quad\quad\quad
  -b  \sqrt{\frac{(l+m)(l-1+m)}{(2l+1)(2l-1)}} |l-1,m-1\rangle 
  \right)\right\} \nonumber \;,
\end{eqnarray}

\begin{eqnarray}
\fl
|c_4(t)\rangle =  \sum_{l} \left\{ f_{n'_+} \exp{(-iE^+_lt)}  \sum_{m} \,
  w_{lm}^{(n)} \left(
  -b \sqrt{\frac{(l+1+m)(l+1-m)}{(2l+1)(2l+3)}}|l+1, m \rangle 
  \right.\right. \nonumber
  \\  \left.\quad\quad\quad\quad\quad\quad\quad\quad\quad
  - a \sqrt{\frac{(l+1+m)(l+2+m)}{(2l+1)(2l+3)}}|l+1, m+1\rangle
  \right) \\ 
    + \,f_{n'_-} \,\exp{(-iE^-_lt)}  \sum_{m} \,  w_{lm}^{(n)}  \left( 
  -b \sqrt{\frac{(l+m)(l-m)}{(2l+1)(2l-1)}}|l-1,m\rangle 
  \right.\nonumber
  \\  \left.\left.\quad\quad\quad\quad\quad\quad\quad\quad\quad
  +a  \sqrt{\frac{(l-m)(l-1-m)}{(2l+1)(2l-1)}} |l-1,m+1\rangle 
  \right)\right\} \nonumber \;.
\end{eqnarray}
Using the explicit form of the WP, one obtains for the average values of 
the spin operators:  

\begin{eqnarray}
\fl
\langle \sigma_x \rangle_t =
2 a b \sum_{lm} \left\{  w_{l,m}^2
  \left[ G_{+} \frac{(l+1)^2-m^2}{(2l+1)^2}
     +G_{-}\frac{l^2-m^2}{(2l+1)^2}
     -F_{+}\frac{(l+1)^2-m^2}{(2l+1)(2l+3)} \right.\right. \nonumber \\
 \hspace{5ex}   \left. -F_{-}\frac{l^2-m^2}{(2l+1)(2l-1)} 
    +2G_{+-}\frac{l(l+1)+m^2}{(2l+1)^2} \cos{(\omega_l t)} \right] 
     \nonumber \\
\fl \hspace{5ex} 
  + w_{l,m}w_{l-2,m-2} \left[ F_{-+} 
\sqrt{\frac{(l+m)(l-1+m)(l-2+m)(l-3+m)}{(2l-1)^2(2l+1)(2l+3)}}
   \cos{(\omega''_lt)} \right] \nonumber  \\
\fl \hspace{5ex}
  + w_{l,m}w_{l,m-2} \frac{\sqrt{(l+m)(l-1+m)}}{(2l+1)}  \hspace{-1ex}\\ 
   \times \left[
   \frac{\sqrt{(l+1-m)(l+2-m)}}{(2l+1)} 
   [G_{+}+G_{-}-2G_{-+}\cos{(\omega_l t)}] \right. \nonumber \\ 
   \left.
   \hspace{3ex}
   -\frac{\sqrt{(l+2-m)(l-3-m)}}{(2l-1)} F_{-}
   -\frac{\sqrt{(l+2-m)(l+1-m)}}{(2l+3)} F_{+} \right] \nonumber \\ 
\fl \hspace{5ex}
  - w_{l,m}w_{l-2,m} \left[2\,  F_{-+} 
  \sqrt{\frac{(l^2-m^2)((l-1)^2-m^2)}{(2l-1)^2(2l+1)(2l-3)}} 
  \cos{(\omega''_l t)} \right] \nonumber \\
\fl \hspace{5ex}  \left.
  + w_{l,m}w_{l+2,m-2} \left[ F'_{-+} 
  \sqrt{\frac{(l+1-m)(l+2-m)(l+3-m)(l+4-m)}{(2l+1)(2l+3)^2(2l+5)}} 
  \cos{(\omega'_l t)} \right] \right\}
\nonumber  ,
\end{eqnarray}

\begin{eqnarray}
\fl
\langle \sigma_y \rangle_t =
2 a b \sum_{lm}  \left\{ w_{l,m}^2 \left[
  \frac{2m}{2l+1} G_{+-} \sin{(\omega_l t)}\right] \right.\\
\fl \hspace{5ex}
  - w_{l,m}w_{l-2,m-2} \left[ F_{-+} 
  \sqrt{\frac{(l+m)(l-1+m)(l-2+m)(l-3+m)}{(2l-1)^2(2l+1)(2l+3)}} 
  \sin{(\omega''_l t)} \right] \nonumber \\
\fl \hspace{5ex}
  + w_{l,m}w_{l-2,m} \left[2\, F_{-+} 
  \sqrt{\frac{(l^2-m^2)((l-1)^2-m^2)}{(2l-1)^2(2l+1)(2l-3)}} 
  \sin{(\omega''_l t)} \right] \nonumber \\
\fl \hspace{5ex}  \left.
  - w_{l,m}w_{l+2,m-2} \left[ F'_{-+} 
  \sqrt{\frac{(l+1+m)(l+2-m)(l+3-m)(l+4-m)}{(2l-1)(2l+3)^2(2l+5)}} 
  \sin{(\omega'_l t)} \right] \right\}  
\nonumber  ,
\end{eqnarray}

and

\begin{eqnarray}
\fl
\langle \sigma_z \rangle_t = \sum_{lm} \left\{ w_{l,m}^2 \left[ 
  a^2 \frac{2m+1}{2l+1} \left( 
  G_{+} \frac{l+1+m}{2l+1} - G_{-} \frac{l-m}{2l+1}
 -F_{+} \frac{l+1+m}{2l+3} + F_{-} \frac{l-m}{2l-1} \right) \right. \right. 
  \nonumber \\ 
  \hspace{2ex} \left.  + b^2 \frac{2m-1}{2l+1}  \left( 
  G_{+} \frac{l+1-m}{2l+1} - G_{-} \frac{l+m}{2l+1}
 -F_{+} \frac{l+1-m}{2l+3} + F_{-} \frac{l+m}{2l-1} \right) \right.
  \nonumber \\ 
  \hspace{2ex} \left. + 4 \,G_{+-}\, \cos{(\omega_l t)} \left(
  a^2 \frac{(l-m)(l+1+m)}{(2l+1)^2} - b^2 \frac{(l+m)(l+1-m)}{(2l+1)^2}
  \right) \right] 
  \nonumber \\ 
\fl \hspace{5ex}  \left. 
  + 4(a^2-b^2)\, w_{l,m}w_{l+2,m}\, F'_{-+} 
  \sqrt{\frac{((l+1)^2-m^2)((l+2)^2-m^2)}{(2l+1)(2l+3)^2(2l+5)}} 
  \cos{(\omega'_l t)} \right\}  
\end{eqnarray}
In the above formulas, the following notations have been introduced:
\begin{eqnarray}
\omega_l &=& (E^+_l-E^-_l)   \;, \\
\omega'_l &=& (E^-_{l+2}-E^+_l)  \;, \\
\omega''_l &=& (E^-_{l}-E^+_{l-2})  
\end{eqnarray}
Note that $\omega'_l=\omega''_{l+2}$. 
Radial integrals are denoted as follows:
\begin{eqnarray}
G_{+} &=& \int_0^{\infty}  \left( g^+_l(r) \right)^2 r^2 dr \;, \\
G_{-} &=& \int_0^{\infty}  \left( g^-_l(r) \right)^2 r^2 dr \;, \\
F_{+} &=& \int_0^{\infty}  \left( f^+_l(r) \right)^2 r^2 dr \;, \\
F_{-} &=& \int_0^{\infty}  \left( f^-_l(r) \right)^2 r^2 dr \;, \\
G_{+-} &=& \int_0^{\infty} g^+_l(r)g^-_l(r)  r^2 dr \;, \\
F_{+-} &=& \int_0^{\infty}  f^+_{l+2}(r) f^-_l(r)  r^2 dr \;, \\
F_{-+} &=& \int_0^{\infty}  f^+_{l-2}(r) f^-_l(r)  r^2 dr \;, \\
F'_{-+} &=& \int_0^{\infty}  f^+_{l}(r) f^-_{l+2}(r)  r^2 dr \;.
\end{eqnarray}
Apart from the case of $G_{+}, G_{-}, F_{+}, F_{-}$ for $l=n-1$,
which are relatively easily obtained analytically, all other radial integrals
have been calculated numerically (using quadruple precision).

The autocorrelation function can be calculated from (\ref{a1}) in a 
straightforward way:
\begin{eqnarray}\label{ac} 
\fl \hspace{4ex}
 \langle \Psi_{r}(0)|\Psi_{r}(t) \rangle 
&=& \sum_l \left\{ \exp{(-iE_l^+t)}\left[ \sum_m w_{l,m}^2 \left(
 a^2\frac{l+1+m}{2l+1}+b^2\frac{l+1-m}{2l+1} \right)\right]\right. 
 \nonumber \\
 && \hspace{2ex} +\left. \exp{(-iE_l^-t)}\left[ \sum_m w_{l,m}^2 \left(
  a^2\frac{l-m}{2l+1}+b^2\frac{l+m}{2l+1} \right)\right]\right\} 
\end{eqnarray}

\section{}\label{aB}

Let us use the following notation:
\begin{equation}\label{b1}
 k = j+1/2 \quad,\quad {\cal E}_k= \frac{E_{nlj}}{m_0c^2} \quad,\quad
 x = (Z\alpha)^2 \,.
\end{equation} 
The exact eigenergies ${\cal E}_k$ (in units ${m_0c^2}$) are given by
\begin{equation}\label{b2}
 {\cal E}_k= \left[ 1+\frac{x^2}{(n-k+\sqrt{k^2-x^2})^2}\right]^{-1/2} \,.
\end{equation} 
Expanding this expression in Taylor series with respect to $x$ one obtains
\begin{equation}\label{b3}
\fl
 {\cal E}_k= 1 - \frac{x^2}{2\,n^2} - \frac{x^4}{4\,n^3}\left(\frac{2}{k} 
 - \frac{3}{2\,n}\right) - \frac{x^6}{4\,n^3}\left( \frac{1}{2\,k^3}
 + \frac{3}{2\,n\,k^2} - \frac{3}{n^2\,k}+\frac{5}{4\, n^3} \right) 
 + [O(x)]^8\,.
\end{equation} 
In eq.~\ref{e8} only the lowest term depending on $k$ in $x^4$ {\em i.e.}
$ \delta_4{\cal E}_k= - \frac{x^4}{4\,n^3}\frac{2}{k}$ has been included. The 
higher order term $ \delta_6{\cal E}_k$ contributes very little, because the
ratio $\delta_6{\cal E}_k/\delta_4{\cal E}_k= x^2(\frac{1}{k^2}+\frac{3}{4nk}
-\frac{3}{2n^2}+\frac{5k}{8n^3})$ reaches the maximum value about 0.0005 for
$\epsilon=0.4$ and $Z=92$ and stays much smaller for lower $Z$.   
Then the time evolution for not too long period is mainly determined by 
the lowest order contribution (\ref{e8}).

The precession time is determined by the derivative 
\begin{equation}\label{b4}
 \frac{\partial {\cal E}_k}{\partial k}\left|_{k=l_{av}}\right. =
  \frac{x^4}{2\,n^3\,k^2}\left[ 1+ \frac{3x^2}{2}\left( \frac{1}{2k^2} +
   \frac{1}{nk} - \frac{1}{2n^2}\right)\right] \;.
\end{equation} 
Again in eqs.~(\ref{e9})-(\ref{e10}) only term of the order of $x^4$ 
has been used. The $x^6$-order term contributes at most about 0.00022
of the $x^4$-order term for $\epsilon=0.4$ and $Z=92$.
Therefore we conclude that the  $x^6$-order term can be safely neglected 
in estimation of the precession time $T_p$.

\end{document}